\documentclass[11pt]{article}

  \usepackage{latexsym}
  \usepackage{amssymb}

  \def\AFOUR{%
  \setlength{\textheight}{9.0in}%
  \setlength{\textwidth}{5.75in}

  \setlength{\topmargin}{-0.375in}%
  \hoffset=-.5in%
  \renewcommand{\baselinestretch}{1.17}%
  \setlength{\parskip}{6pt plus 2pt}%
  }

  \AFOUR                                          

  \makeatletter
  \def\section{\@startsection {section}{1}{\z@}{-3.5ex plus -1ex minus
   -.2ex}{2.3ex plus .2ex}{\large\bf}}
  \def\subsection{\@startsection{subsection}{2}{\z@}{-3.25ex plus -1ex minus
   -.2ex}{1.5ex plus .2ex}{\normalsize\bf}}
  \makeatother

  \makeatletter
  \@addtoreset{equation}{section}
  
  \makeatother
\newtheorem{theorem}{Theorem}[section]
\newtheorem{proposition}[theorem]{Proposition}

\newtheorem{corollary}[theorem]{Corollary}

\begin{document}
\begin{titlepage}

\begin{center}

\vspace{1cm}

\vskip .5in
 {\Large\bf $(q,\mu)$ and $(p,q,\zeta)-$exponential functions:
 Rogers-Szeg\H{o}
polynomials and Fourier-Gauss transform}

\vspace{10pt}

{\bf   M. N. Hounkonnou and E. B. Ngompe Nkouankam}

\vspace{10pt}

\vspace{0.5cm}

{\em International Chair of Mathematical Physics
and Applications} \\
{\em ICMPA-UNESCO Chair}\\
{\em Universit\'e d'Abomey-Calavi}\\
{\em 072 B.P. 50 Cotonou, Republic of Benin}\\
E-mail:
norbert.hounkonnou@cipma.uac.bj\footnote{With
copy to hounkonnou@yahoo.fr}

\vspace{1.0cm}

\today

\begin{abstract}
From the realization of $q-$oscillator algebra in terms of
generalized derivative, we compute  the matrix elements from
deformed  exponential functions and deduce generating functions
associated with  Rogers-Szeg\H{o} polynomials as well as
 their
relevant properties. We also compute the
matrix elements  associated to
the $(p,q)-$oscillator algebra (a generalization
of the $q-$one) and  perform the Fourier-Gauss transform
of a generalization of the
deformed exponential functions.
\end{abstract}
\end{center}


Pacs numbers: 02.30.Gp, 02.20.Uw

\end{titlepage}
\makeatother


 \section{Introduction}
The deformation of quantum algebras, and consequently quantum
$q$-deformations of Lie algebras and related properties
 still continue to be of  great relevance  in mathematics and
 physics due to their important applications
 in the quantum field theories  and
quantum groups
 \cite{burban}.

Lie group theory and representations  spawned new ideas and results
 providing a unifying framework
  for discussing special
functions. The latter  appear as solutions of differential equations
describing specific physical problems and satisfy suitable
properties such as orthogonality giving rise to  generalized Fourier
analysis. Specifically, the development of the theory of group
representations has made it possible to comprehend the theory of the
most important classes of special functions from a single point of
view. Indeed, the appraisal of the importance of individual classes
of special functions has greatly changed during the last hundred
years. In particular, the class of special functions associated with
the hypergeometric function and its various special and degenerate
cases, the functions of Bessel and Legendre, the orthogonal
polynomials of Jacobi, Tchebychev, Laguerre, Hermite, etc. which
play a big role in different branches of mathematics and its
numerous applications to astronomy and mathematical physics, lends
itself to a group-theoretical treatment.

The connection between special functions and group representation
was first discovered by  Cartan \cite{cartan} in 1929. However, a
connection between the theory of special functions and the theory of
invariants, which is one of the aspects of the theory of group
representations, was established even earlier. The application of
the theory of representations to quantum mechanics played a
significant part in the investigation of these  connections. Further
development in this field was stimulated by the works of
Gel'fand and  Naimark and their students and collaborators in
the field of infinite-dimensional group representations. They linked
the theory of group representations with the automorphic functions,
and developed the theory of special functions over finite fields,
investigated the special functions in homogeneous domains, and so
on. A more detailed statement of the above, including a systematic
account of the theory of special functions from the group
theoretical point of view, is found in form of a nice compilation in
the book by  Vilenkin \cite{vilenkin}, the most used classics
in the field.

In the same vein, the discovery  of quantum groups has in turn
prompted the undertaking of a systematic investigation of the
algebraic
 properties of the $q-$analogs
of  special functions.  The matrix elements in representation
spaces, yielding relevant properties of $q-$special functions,  are
defined from deformed exponential functions of the generators
associated with a given deformed algebra. Intense
 research activities in such an
area as $q$-special functions are mainly motivated by their
importance  in quantum theory. This work highlights some relevant
properties of Rogers-Szeg\H{o} polynomials from $(q,\mu)-$
exponential functions and provides with
matrix elements and
Fourier-Gauss transform
of $(p,q,\mu,\nu)-$exponential functions.

The paper is organized as follows. In section 2, we give some
 well-known connection between quantum algebra representations and
$q-$polynomials.  In section 3,  we  compute the matrix elements of
a known deformed exponentials  related to the generators of standard
$q-$oscillator \cite{arik}, which we  use  to deduce some properties
  of   Rogers-Szeg\H{o}
polynomials.  We then
provide a generalization of the introduced deformed exponential
functions and study the associated
matrix elements and the  Fourier-Gauss
transform in section 4.

\section{Overview of known results}
In this section, let us briefly recall the most popular matrix
elements issued from irreducible quantum algebra representations
generating $q-$polynomials.

\begin{enumerate}
\item[$(1)$]{\it Little $q-$Jacobi and $q-$Legendre polynomials}.
They arise from the noncommutative algebra
   $U_{q}(sl_{2}(\mathbb{C}))$
generated by $I$ and $X_{\pm}$, $q^{\pm \frac{H}{2}}$ with the
relations  \cite{majid} $q^{\pm \frac{H}{2}}q^{\mp \frac{H}{2}}=I$
and
\begin{equation}
q^{\frac{H}{2}}X_{\pm}q^{-\frac{H}{2}}=q^{\pm}X_{\pm}
\qquad [X_{+},X_{-}]=\frac{q^{H}-q^{-H}}{q-q^{-1}}
\end{equation}
and $V_{\epsilon}(\lambda)$,  the irreducible highest weight
$U_{\epsilon}(sl_{2}(\mathbb{C}))-$module with highest weight
$\lambda \in \mathbb{N}$. Indeed, as quoted in \cite{jacobi}, the
matrix elements $C_{\nu;\mu}^{\lambda}$
 of the irreducible
$U_{\epsilon}(sl_{2}(\mathbb{C}))-$module $V_{\epsilon}(\lambda)$,
with respect to an orthonormal basis are related
to little  $q-$Jacobi polynomials
\begin{eqnarray}
p_{n}(z;\alpha,\beta;q)=\;
_{2}\phi_{1}
\left(
\begin{array}{cc}
q^{-n},q^{n+1}\alpha\beta& \\
&;q,qz\\
\alpha q&
\end{array}
\right).
\end{eqnarray}
and defined by the following statement.
\begin{proposition}
For different values of $\mu$ and $\nu$, the matrix elements
$C_{\nu;\mu}^{\lambda}$ of the irreducible
$U_{\epsilon}(sl_{2}(\mathbb{C}))-$module
of highest weight $\lambda$ are given as follows:
\begin{itemize}
\item[$(i)$] if $\mu+\nu\geq 0$ and $\mu\geq\nu$,
\begin{eqnarray}
 C_{\nu;\mu}^{\lambda}
&=&a^{(\mu+\nu)/2}c^{(\mu-\nu)/2}q^{(\lambda-\mu)(\nu-\mu)/4}
\left(\left[
\begin{array}{c}
\frac{1}{2}(\lambda-\nu)\\
\frac{1}{2}(\mu-\nu)
\end{array}
\right]_{q^{2}}^{P}\right)^{1/2}
\cr
&& \times
\left(\left[
\begin{array}{c}
\frac{1}{2}(\lambda+\mu)\\
\frac{1}{2}(\mu-\nu)
\end{array}
\right]_{q^{2}}^{P}\right)^{1/2} p_{\frac{1}{2}(\lambda-\nu)}
(-q^{-1}bc;q^{\mu-\nu},q^{\mu+\nu};q^{2})
\end{eqnarray}
\item[$(ii)$]
if $\mu+\nu\geq 0$ and $\mu\leq\nu$,
\begin{eqnarray}
 C_{\nu;\mu}^{\lambda}
&=&a^{(\mu+\nu)/2}b^{(\nu-\mu)/2}q^{(\lambda-\nu)(\mu-\nu)/4}
\left(\left[
\begin{array}{c}
\frac{1}{2}(\lambda+\mu)\\
\frac{1}{2}(\nu-\mu)
\end{array}
\right]_{q^{2}}^{P}\right)^{1/2}\cr
&&\times
\left(\left[
\begin{array}{c}
\frac{1}{2}(\lambda-\mu)\\
\frac{1}{2}(\nu-\mu)
\end{array}
\right]_{q^{2}}^{P}\right)^{1/2} p_{\frac{1}{2}(\lambda-\nu)}
(-q^{-1}bc;q^{\nu-\mu},q^{\mu+\nu};q^{2})
\end{eqnarray}
\item[$(iii)$] if $\mu+\nu\leq 0$ and $\mu\leq\nu$,
\begin{eqnarray}
 C_{\nu;\mu}^{\lambda}
=q^{-(\lambda+\mu)(\nu-\mu)/4}
\left(\left[
\begin{array}{c}
\frac{1}{2}(\lambda+\nu)\\
\frac{1}{2}(\nu-\mu)
\end{array}
\right]_{q^{2}}^{P}\right)^{1/2}
\left(\left[
\begin{array}{c}
\frac{1}{2}(\lambda-\mu)\\
\frac{1}{2}(\nu-\mu)
\end{array}
\right]_{q^{2}}^{P}\right)^{1/2}\cr
\qquad\qquad \times p_{\frac{1}{2}(\lambda+\nu)}
(-q^{-1}bc;q^{\mu-\nu},q^{-\mu-\nu};q^{2})b^{(\nu-\mu)/2}d^{-(\mu+\nu)/2}
\end{eqnarray}
\item[$(iv)$] if $\mu+\nu\leq 0$ and $\mu\geq\nu$,
\begin{eqnarray}
 C_{\nu;\mu}^{\lambda}
=q^{-(\lambda+\nu)(\mu-\nu)/4}
\left(\left[
\begin{array}{c}
\frac{1}{2}(\lambda-\nu)\\
\frac{1}{2}(\mu-\nu)
\end{array}
\right]_{q^{2}}^{P}\right)^{1/2}
\left(\left[
\begin{array}{c}
\frac{1}{2}(\lambda+\mu)\\
\frac{1}{2}(\mu-\nu)
\end{array}
\right]_{q^{2}}^{P}\right)^{1/2}\cr
\qquad\qquad \times p_{\frac{1}{2}(\lambda+\nu)}
(-q^{-1}bc;q^{\mu-\nu},q^{-\mu-\nu};q^{2})b^{(\mu-\nu)/2}d^{-(\mu+\nu)/2}
\end{eqnarray}
where
\begin{eqnarray}
\left[
\begin{array}{c}
m\\
n
\end{array}
\right]_{q}^{P}&=
&\frac{[m]_{q}^{P}!}{[n]_{q}^{P}![m-n]_{q}^{P}!}\cr
[n]_{q}^{P}!&=&
[n]_{q}^{P}[n-1]_{q}^{P}\ldots[2]_{q}^{P}
[1]_{q}^{P}\cr
[n]_{q}^{P}&=&\frac{q^{n}-q^{-n}}{q-q^{-1}}.
\end{eqnarray}
\item[$(v)$]For $\lambda$  even and
$\mu=\nu=0$.  one obtains
\begin{equation}
C_{00}^{\lambda}=p_{\lambda/2}(-q^{-1}bc;1,1;q^{2})
\end{equation}
which is a little $q-$Legendre polynomial with argument $-q^{-1}bc$.
\end{itemize}
\end{proposition}
The $q$-hypergeometric series
$_r\phi_s$  are defined by
\begin{eqnarray}
&& _r\phi_s
\left(
\begin{array}{cc}
a_{1}, a_{2}, \ldots, a_{r}&\\
&;q,x\\
b_{1},b_{2},\ldots, b_{s} &
\end{array}
\right)\\
&&=\sum_{k=0}^{+\infty}
\frac{(a_{1};q)_{k}\ldots(a_{r};q)_{k}}{
(q;q)_{k}(b_{1};q)_{k}\ldots(b_{s};q)_{k}}
[(-1)^{k}q^{k(k-1)/2}]^{1+s-r}x^{k}
\end{eqnarray}
where $(a;q)_{k}=
(1-a)(1-aq)\ldots(1-aq^{k-1}).$
\item[$(2)$]{\it Hahn-Exton $q-$Bessel functions.}
They can be generated considering the quantized universal enveloping,
$U_{\epsilon}(\frak{e}_{2})$, of $E(2)$ through the following statements.

\begin{theorem}
Every irreducible unitarizable representation of
$U_{\epsilon}(\frak{e}_{2})$ on which  the
 $*-$structure
$L$ acts semi-simply  with  finite-dimensional eigenspaces is
equivalent to the representation
 $\rho_{\pm}^{\lambda,\mu}$
 on the Hilbert space
$l^{2}(\mathbb{Z})$ given by
\begin{eqnarray}
\rho_{\pm}^{\lambda,\mu}(L)(e_{k})&=&
\pm\mu\epsilon^{k}e_{k}\cr
\rho_{\pm}^{\lambda,\mu}(X^{+})(e_{k})&=&\lambda\mu\epsilon^{k+1}
e_{k+1}\cr
\rho_{\pm}^{\lambda,\mu}(X^{-})(e_{k})&=& \pm\bar{\lambda}
\mu^{-1}\epsilon^{1-k}e_{k-1}
\end{eqnarray}
in terms of the standard orthonormal basis $\{e_{k}\}_{k\in\mathbb{Z}}$ of
$l^{2}(\mathbb{Z})$,
where $L^{\pm 1}$, $X^{\pm}$ are the generators of $U_{\epsilon}(\frak{e}_{2})$;
 $\lambda,\mu\in\mathbb{R}$ with $\lambda>0$ and $1\leq|\mu|<\epsilon$.
\end{theorem}

The matrix elements of $\rho_{+}^{\lambda, 1}$ and
$\rho_{+}^{\lambda,\epsilon^{1/2}}$ involve the
Hahn-Exton $q-$Bessel functions
\begin{equation}
J_{n}(z;q)=z^{n}\frac{(q^{n+1};q)_{\infty}}{(q;q)_{\infty}}\;
_{1}\phi_{1}
\left(
\begin{array}{cc}
0& \\
 &;q,qz^{2}\\
q^{n+1}&
\end{array}
\right).
\end{equation}
\begin{proposition}
\cite{bessel}
The following formulas give the matrix elements of the representations
$\rho_{+}^{\lambda,1}$  and $\rho_{+}^{\lambda,\epsilon^{1/2}}$:
\begin{itemize}
\item[$(i)$]
for $j\geq k$,
\begin{eqnarray}
 (\rho_{+}^{\lambda,1})_{jk}&=&
 \left(\frac{\lambda(1-q^{2})}{q^{k+1}}\right)^{j-k}
 \frac{a^{j+k}b^{j-k}}{(q^{2};q^{2})_{j-k}}\cr
&&\times
_{1}\phi_{1}
\left(
\begin{array}{cc}
0& \\
 &;q^{2},-(1-q^{2})^{2}\lambda^{2}q^{-2k-1}bc\\
q^{2(j-k+1)}&
\end{array}
\right) \cr
 (\rho_{+}^{\lambda,\epsilon^{1/2}})_{jk}&=&
 \left(\frac{\lambda(1-q^{2})}{q^{k+\frac{3}{2}}}\right)^{j-k}
 \frac{a^{j+k+1}b^{j-k}}{(q^{2};q^{2})_{j-k}}\cr
&&\times
_{1}\phi_{1}
\left(
\begin{array}{cc}
0& \\
 &;q^{2},-(1-q^{2})^{2}\lambda^{2}q^{-2k-2}bc\\
q^{2(j-k+1)}&
\end{array}
\right)
 \end{eqnarray}
\item[$(ii)$]
for $j\leq k$,
\begin{eqnarray}
 (\rho_{+}^{\lambda,1})_{jk}&=&
 \left(\frac{\lambda(1-q^{2})}{q^{j+1}}\right)^{k-j}
 \frac{a^{j+k}c^{k-j}}{(q^{2};q^{2})_{k-j}}\cr
&&\times
_{1}\phi_{1}
\left(
\begin{array}{cc}
0& \\
 &;q^{2},-(1-q^{2})^{2}\lambda^{2}q^{-2j-1}bc\\
q^{2(k-j+1)}&
\end{array}
\right) \cr
 (\rho_{+}^{\lambda,\epsilon^{1/2}})_{jk}&=&
 \left(\frac{\lambda(1-q^{2})}{q^{j+\frac{3}{2}}}\right)^{k-j}
 \frac{a^{j+k+1}c^{k-j}}{(q^{2};q^{2})_{k-j}}\cr
&&\times
_{1}\phi_{1}
\left(
\begin{array}{cc}
0& \\
 &;q^{2},-(1-q^{2})^{2}\lambda^{2}q^{-2j-2}bc\\
q^{2(k-j+1)}&
\end{array}
\right).
 \end{eqnarray}
 \end{itemize}
 \end{proposition}
 \item[$(3)$]{\it Big $q-$Jacobi polynomials.}
 They can be built using  the connection between  big $q-$Jacobi polynomials
 and quantum spheres.

 \begin{proposition}
 The quantized algebra $\mathcal{F}_{\epsilon}(S^{2})$ of functions on the
 $2-$spheres is isomorphic to the associative algebra over
 $\mathbb{C}$ with generators $x$, $y$, $z$ satisfying  the relations
\begin{eqnarray}
 xz=\epsilon^{-2}zx \qquad yz=\epsilon^{2}zy\\
 xy=-\epsilon^{-1}z(1-\epsilon^{-2}z)\qquad yx=-\epsilon z(1-z)
 \end{eqnarray}
and the $*-$structure
 \begin{eqnarray}
 x^{*}=-\epsilon^{-1}y \qquad y^{*}=-\epsilon x\qquad z^{*}=z.
 \end{eqnarray}
 The monomials $y^{j}z^{k}x^{l}$ with $j$, $k$, $l$ $\in\mathbb{N}$, are
 basis  of $\mathcal{F}_{\epsilon}(S^{2})$ over $\mathbb{C}$. Moreover,
 $\mathcal{F}_{\epsilon}(S^{2})$ is a quantum $sl_{2}(\mathbb{C})-$space.
 \end{proposition}

  The algebra of polynomial
 functions on the classical
 $2-$spheres, as representation of $SO_{3}$, contains each irreducible
 representation
 exactly once \cite{vilenkin}. The same result holds in the quantum case \cite{book}:
 \begin{proposition}
  As an $\mathcal{F}_{\epsilon}(S^{2})-$comodule, $\mathcal{F}_{\epsilon}(S^{2})$
  decomposes  into irreducibles as follows:
  \begin{equation}
  \mathcal{F}_{\epsilon}(S^{2})\cong
  \bigoplus_{\lambda\in 2\mathbb{N}}V_{\epsilon}
  (\lambda)
  \end{equation}
  where $V_{\epsilon}(\lambda)$ is the irreducible
   $\mathcal{F}_{\epsilon}(S^{2})-$comodule of dimension $\lambda + 1$.
  \end{proposition}
 From this proposition, there exist a unique basis
  $\{S^{\lambda}_{\mu}\}$ of $\mathcal{F}_{\epsilon}(S^{2})$, where
  $\mu=\lambda, \lambda-2,\ldots,-\lambda$, $\lambda\in 2\mathbb{N}$ such that
  $S^{0}_{0}=1$  and
  \begin{equation}
  \Delta_{\mathcal{F}_{\epsilon}}(S_{\mu}^{\lambda})=\sum_{\nu}
  C^{\lambda}_{\mu;\nu}\bigotimes S_{\mu}^{\lambda}
  \end{equation}
 where $\Delta_{\mathcal{F}_{\epsilon}}$ is the comultiplication defined
 by
\begin{equation}
\Delta_{\mathcal{F}_{\epsilon}}:
\mathcal{F}_{\epsilon}(S^{2})\longrightarrow
\mathcal{F}_{\epsilon}(S^{2})\bigotimes\mathcal{F}_{\epsilon}(S^{2}).
\end{equation}
 The $S^{\lambda}_{\mu}$ are $q-$analogues  of spherical  functions.
 They can be  derived in terms of big $q-$Jacobi polynomials
 \begin{eqnarray}
 P_{n}(z;\alpha,\beta;q)=\;
 _{3}\phi_{2}
\left(
\begin{array}{cc}
q^{-n},q^{n+1}\alpha\beta,q\alpha z& \\
 &;q,q\\
q\alpha,0&
\end{array}
\right).
 \end{eqnarray}
 by the following statement \cite{noumi}
  \begin{proposition}
 The $q-$spherical functions $S^{\lambda}_{\mu}$
 are given by the following
 formulas:
 \begin{itemize}
 \item[$(i)$]  if $\mu\geq 0$,
 \begin{eqnarray}
S^{\lambda}_{\mu} =
  (-1)^{(\lambda-\mu)/2}q^{-(\lambda-\mu)(\lambda+3\mu+6)/8}
 \left(\left[
\begin{array}{c}
\lambda\\
\frac{1}{2}(\lambda-\mu)
\end{array}
\right]_{q^{2}}^{P}\right)^{-1/2}\cr
\qquad\qquad\qquad
\times
\left[
\begin{array}{c}
\frac{1}{2}\lambda\\
\frac{1}{2}(\lambda-\mu)
\end{array}
\right]_{q^{2}}^{P}
\times y^{\mu/2}P_{\frac{1}{2}(\lambda-\mu)}(z;q^{\mu},q^{\mu};q^{2})
 \end{eqnarray}
\item[$(ii)$]  if $\mu\leq 0$,
 \begin{eqnarray}
 S^{\lambda}_{\mu} =
  (-1)^{(\lambda+\mu)/2}q^{-(\lambda+\mu)(\lambda-3\mu+6)/8}
  \left(\left[
\begin{array}{c}
\lambda\\
\frac{1}{2}(\lambda+\mu)
\end{array}
\right]_{q^{2}}^{P}\right)^{-1/2}\cr
\qquad\qquad\qquad
\times\left[
\begin{array}{c}
\frac{1}{2}\lambda\\
\frac{1}{2}(\lambda+\mu)
\end{array}
\right]_{q^{2}}^{P}
\times x^{-\mu/2}P_{\frac{1}{2}(\lambda+\mu)}(z;q^{-\mu},q^{-\mu};q^{2}).
 \end{eqnarray}
 \end{itemize}
 \end{proposition}
 \item[$(4)$]{\it $q-$analogs of Bessel functions.}  Floreanini  \cite{ref1}
 showed that the  matrix elements of the two dimensional quantum Euclidian
algebra $\mathcal U_q(E(2))$ representations are given by
\begin{equation}
 U_{k,n}(\alpha,\beta)=q^{(k-n)^{2}/2}\left(-\frac{\alpha}{\beta}
\right)^{(k-n)/2}J^{(2)}_{k-n}\left(
2w(-\frac{\alpha\beta}{q})^{1/2};q\right)
\end{equation}
 where $n, k\in\mathbb{Z}$ and
  $J^{(2)}_{\nu}(x;q)$ are  $q-$analogs of Bessel
functions  defined by
\begin{equation}
J^{(2)}_{\nu}(x;q)=\sum_{n=0}^{+\infty}q^{n(n+\nu)}
\frac{(-1)^{n}}{(q;q)_{n}(q;q)_{n+\nu}}\left(\frac{x}{2}\right)^{2n+\nu}.
\end{equation}
The authors  used this  model  to entail a $q$-analog of Graf's
addition formula for Bessel functions and $q-$analog of the
Fourier-Gegenbauer expansion.
\end{enumerate}
 In the same vein, a  series of papers
\cite{flo2}-\cite{koorwi4}
should be quoted for their relevance to
$q-$orthogonal polynomials and $q-$special functions deduced from
matrix elements or basis vectors of quantum algebra representations.

In the next section, we aim at  using this formalism to investigate
   some properties  of   Rogers-Szeg\H{o}
polynomials which play an important role in the theory of orthogonal
polynomials, particularly in the study of the Askey-Wilson
polynomials \cite{h1,h2,h3}.

\section{$(q,\mu)$-exponential functions:
 matrix elements}

 The $q-$oscillator algebra is generated  by
 three elements $A_{-}$,
  $A_{+}$ and $N$ obeying the relations
   \cite{arik}

\begin{eqnarray}
A_{-}A_{+}-A_{+}A_{-}=q^{N}\qquad
A_{-}A_{+}-qA_{+}A_{-}=I
\label{alg11}\\
\;[N,A_{-}]=-A_{-}\qquad [N,A_{+}]=A_{+}.
\label{alg22}
\end{eqnarray}

In general, the  parameter $q$  may be real or a
phase factor. Throughout, we suppose that it is
  real and positive.

The algebra (\ref{alg11})-(\ref{alg22}) admit a class
 of irreducible representattions defined on the
 vector space spanned by the basis vectors
 $\theta_{n}$, $n = 0, 1, 2, . . .$
 such that \cite{galetti}

\begin{eqnarray}
\label{ge}
A_{+}\theta_{n}&=&\theta_{n+1}\label{ge1}\\
A_{-}\theta_{n}&=& \frac{1-q^{n}}{1-q}
\theta_{n-1}\label{ge2}.
\end{eqnarray}

One can easily verify that these definitions are
compatible with the relations
(\ref{alg11})-(\ref{alg22}).

\subsection{Matrix elements}

In order to make a link between the algebra
(\ref{alg11})-(\ref{alg22}) and $q-$polynomials,
one can note that the representation
(\ref{ge1})-(\ref{ge2}) is given in terms of
the Rogers-Szeg\H{o}
polynomials
\begin{equation}
\theta_{n}(y)
\equiv{\cal H}_{n}(y|q)=\sum_{k=0}^{n}
\left[
\begin{array}{c}
n\\
k
\end{array}
\right]_{q}y^{k}
\end{equation}
with the $q$-binomial coefficients given by
\begin{equation}
\left[
\begin{array}{c}
n\\
k
\end{array}
\right]_{q}=
 \frac{(q;q)_{n}}{(q;q)_{k}(q;q)_{n-k}}.
  \end{equation}
Indeed, by taking $A_{+}$, $A_{-}$ to be
the following operators, defined in terms
of the $q-$Jackson derivative  acting on
the space of analytic functions

\begin{eqnarray}
A_{-}f(y)&:=& _{q}D_{y}f(y)\\
A_{+}f(y)&:=& (1+y)f(y)-(1-q)y _{q}D_{y}f(y)
\end{eqnarray}

where
\begin{equation}
_{q}D_{y}f(y)=\frac{f(y)-f(qy)}{(1-q)y}
\end{equation}
one proves that relations
(\ref{ge1})-(\ref{ge2}) are verified.

In order to reproduce the exponential mapping,
necessary to pass  from Lie algebras
 to Lie groups, we  consider the following
  $(q,\mu)-$exponential function
 \cite{flore}
\begin{equation}
\label{gene1}
E_{q}^{(\mu)}(z)
=\sum_{n=0}^{+\infty}
\frac{q^{\mu n^{2}}}{(q;q)_{n}}z^{n}
\qquad \mu \geq 0\qquad 0<q<1.
\end{equation}
In the limit $q \to 1$,
$E_{q}^{(\mu)}((1-q)z)$ tend to the
ordinary exponential: $\lim_{q\to 1}
 E_{q}^{(\mu)}((1-q)z)= \exp(z)$.
 We also note that
 for some specific values of $\mu$, they
correspond to standard $q$-exponentials.
Indeed, for $\mu=0$ and $\mu = 1/2$ one has
 \cite{koek}

\begin{eqnarray}
\label{num1}
e_{q}(z)\equiv E^{(0)}(z)=
 \sum_{n=0}^{+\infty}\frac{1}{(q;q)_{n}}z^{n}
=\frac{1}{(z;q)_{\infty}}
\end{eqnarray}
\begin{eqnarray}
 \label{num2}
E^{(1/2)}(z)&=& E_{q}(q^{1/2}z)=
(-q^{1/2}z;q)_{\infty}.
\end{eqnarray}

For an algebraic interpretation of special
class of $q$-hypergeometric  functions,
let us  introduce the following operators
\begin{equation}
\mathcal{U}^{(\mu,\nu)}(\alpha,\beta)
=E_{q}^{(\mu)}(\alpha(1-q)A_{+})
E_{q}^{(\nu)}(\beta(1-q)A_{-}).
\end{equation}
 Then,  in the limit
$q \to 1$, they go into the Lie group element
$\exp(\alpha
A_{+})\exp(\beta A_{-}).$
 Their matrix elements, in the
representation space spanned by the vectors
 $\theta_{n}$,
are defined by
\begin{equation}
\label{matel}
 \mathcal{U}^{(\mu,\nu)}(\alpha,\beta)\theta_{n}
 =\sum_{m=0}^{+\infty}U_{m,n}^{\mu,\nu}
 (\alpha,\beta)
 \theta_{m}
\end{equation}
and, when evaluated, are found to involve
 generalized hypergeometric
 functions.
 By using  (\ref{gene1}) and various identities
  for the
  $q-$shifted  factorials, one explicitly finds

\begin{eqnarray}
 U_{m,n}^{\mu,\nu}(\alpha,\beta)&=&
 \beta^{n-m}
 \left[
\begin{array}{c}
n\\
m
\end{array}
 \right]_{q}q^{\nu(n-m)^{2}}\cr
 &&\quad \times \mathcal{Q}^{(\mu,\nu)}_{m}
 \left(-\alpha\beta(1-q);q^{n-m}|q\right)
 \quad {\rm if} \quad n\geq m
 \label{mat1}
\end{eqnarray}
\begin{eqnarray}
U_{m,n}^{\mu,\nu}(\alpha,\beta)&=&
\frac{\left[(1-q)\alpha\right]^{m-n}
q^{\mu(m-n)^{2}}}{(q;q)_{m-n}}\cr
&& \quad \times \mathcal{Q}^{(\nu,\mu)}_{n}
 \left(-\alpha\beta(1-q);q^{m-n}|q\right)
 \quad {\rm if} \quad  m\geq n
 \label{mat2}
\end{eqnarray}

where $\mathcal{Q}^{(\mu,\nu)}_{n}
 \left(x;q^{\gamma}|q\right)$  is the
  polynomial given  by
\begin{equation}
 \mathcal{Q}^{(\mu,\nu)}_{n}
 \left(x;q^{\gamma}|q\right)=\sum_{k=0}^{n}
 \frac{q^{k^{2}(\mu+\nu) + (2\nu\gamma+n)k}
 (q^{-n};q)_{k}}
 {(q;q)_{k}(q^{\gamma+1};q)_{k}}q^{-k(k-1)/2}
 x^{k}.
\end{equation}
Note that in passing from expression
(\ref{mat1}) to (\ref{mat2})
for $U_{m,n}^{\mu,\nu}(\alpha,\beta)$
 or vice-versa, $m$ and $n$ as
well as $\mu$ and $\nu$ are exchanged in
 the polynomials $
\mathcal{Q}^{(\mu,\nu)}_{n}
\left(x;q^{\gamma}|q\right)$.

The connection with standard
$q$-polynomials is observed for particular
values of $\mu$ and $\nu$.

For instance, for $\mu=\nu=0$
\begin{eqnarray}
\mathcal{Q}^{(0,0)}_{n}
 \left(x;q^{\gamma}|q\right)= \; _{3}\phi_{1}
\left(
\begin{array}{cc}
q^{-n},0,0&\\
&;q,-xq^{n}\\
q^{\gamma+1}&
\end{array}
\right)
\end{eqnarray}
for $\mu=0,$  $\nu=\frac{1}{2}$
\begin{eqnarray}
\mathcal{Q}^{(0,1/2)}_{n}
 \left(x;q^{\gamma}|q\right)&=& \; _{2}\phi_{1}
\left(
\begin{array}{cc}
q^{-n},0& \\
&;q,xq^{\gamma+n+\frac{1}{2}}\\
q^{\gamma+1}&
\end{array}
\right)  \cr
&=& p_{n}\left(xq^{\gamma+n-\frac{1}{2}};
q^{\gamma},0|q\right)
\end{eqnarray}
where $p_{n}(z;\alpha,\beta|q)$ is the little
 $q-$Jacobi polynomials,
 for $\mu=\nu=\frac{1}{2}$
\begin{eqnarray}
\mathcal{Q}^{(1/2,1/2)}_{n}
 \left(x;q^{\gamma}|q\right)&=&\; _{1}\phi_{1}
\left(
\begin{array}{cc}
q^{-n}& \\
&;q,xq^{\gamma+n+1}\\
q^{\gamma+1}&
\end{array}
\right)\\
&=& \frac{(q;q)_{n}}{(q^{\gamma+1};q)_{n}}
L_{n}^{(\gamma)}(x)
\end{eqnarray}
where $L_{n}^{(\gamma)}(x)$ are the
 $q-$Laguerre polynomials \cite{koek}.

\subsection{Main properties of
 Rogers-Szeg\H{o}
 polynomials}

Let us turn back  to the relations
(\ref{ge1})-(\ref{ge2}) and make
use of the matrix elements  to derive
 some properties  of
 the Rogers-Szeg\H{o} polynomials.

\begin{theorem}
 The  Rogers-Szeg\H{o} polynomials possess
  the
 generating functions
 \begin{itemize}
 \item[$i)$]
 \begin{equation}
\label{generate}
\mathcal{S}_{q}(\alpha;y)\equiv
\frac{1}{(\alpha;q)_{\infty}
(\alpha y;q)_{\infty}}
=\sum_{m=0}^{+\infty}
\frac{\alpha^{m}}{(q;q)_{m}}
{\cal H}_{m}(y|q)
\end{equation}
  \item[$ii)$]
  \begin{equation}
  \label{generatex2}
_{1}\phi_{1}\left(
\begin{array}{cc}
0&\\
&;q,ty\\
-q^{1/2}t&
\end{array}
\right)(-tq^{1/2};q)_{\infty}=
\sum_{m=0}^{+\infty}
\frac{t^{m}q^{m(m-1)/2}}{(q;q)_{m}}
{\cal H}_{m}(y|q).
\end{equation}
 \end{itemize}
\end{theorem}

{\bf Proof.}

From the definition of the matrix elements and
(\ref{mat2}) one can write
\begin{eqnarray}
\label{case}
\mathcal{U}^{(\mu,0)}(\alpha/(1-q),0)\cdot
1 &=& \sum_{m=0}^{+\infty}
\frac{\alpha^{m}q^{\mu m^{2}}}{(q;q)_{m}}
{\cal H}_{m}(y|q).
\end{eqnarray}
This  relation indicates  that if
$\mathcal{U}^{(\mu,0)}(\alpha/(1-q),0)\cdot 1$
can be write in a closed form then, this
would be generating  functions for  the
Rogers-Szeg\H{o} polynomials.  Let us consider
two particular cases.

 $i)$ For  $\mu=0$, (\ref{case}) becomes
\begin{equation}
\label{case11}
E^{(0)}_{q}(\alpha A_{+})\cdot 1=
\sum_{k,m=0}^{+\infty}\frac{\alpha^{m}}
{(q;q)_{m-k}(q;q)_{k}}
y^{k}.
\end{equation}
By introducing the new summation index
$l=m-k$ on the right-hand side of (\ref{case11}),
one obtains
\begin{equation}
E^{(0)}_{q}(\alpha A_{+})\cdot 1
=e_{q}(\alpha)e_{q}(\alpha y).
\end{equation}
Hence,  the following  generating
relation
\begin{equation}
\label{generatex1}
\frac{1}{(\alpha;q)_{\infty}
(\alpha y;q)_{\infty}}
=\sum_{m=0}^{+\infty}
\frac{\alpha^{m}}{(q;q)_{m}}{\cal H}_{m}(y|q).
\end{equation}

 $ii)$ For $\mu=1/2$,
 (\ref{case}) can be rewritten as
\begin{equation}
\label{case1}
\mathcal{U}^{(1/2,0)}(\alpha/(1-q),0)\cdot
1 =
\sum_{m,k=0}^{+\infty}\frac{q^{m(m-1)/2}}
{(q;q)_{m-k}(q;q)_{k}}
(\alpha q^{1/2})^{m}y^{k}.
\end{equation}
  The two sums are reorganized by using
  $l=m-k$  instead
of  $m$ as summation index. This allows us
 to perform the sum over
$l$ thanks to the explicit expansion for
(\ref{num2}) and the
Heine's binomial theorem which states that
 \cite{koek}
\begin{equation}
\sum_{n=0}^{+\infty}\frac{(a;q)_{n}}
{(q;q)_{n}}z^{n}=\frac{(az;q)_{\infty}}
{(z;q)_{\infty}}.
\end{equation}
One obtains
\begin{equation}
\label{case2}
\mathcal{U}^{(1/2,0)}(\alpha/(1-q),0)\cdot 1 =\;
_{1}\phi_{1}\left(
\begin{array}{cc}
0&\\
&;q,\alpha q^{1/2}y\\
-\alpha q&
\end{array}
\right)(-\alpha q;q)_{\infty}.
\end{equation}
Finally, we set $t=\alpha q^{1/2}$  on the
right-hand sides of (\ref{case1}) and
  (\ref{case2}) to find the generating function
 identity (\ref{case}). $\Box$

It is worth mentioning that the formula
(\ref{generate}) which was
here constructively derived using algebraic
 methods, coincide with
the one given in \cite{mouad}.

The  generating function (\ref{generate})
 can be used to  determine
another realization of Rogers-Szeg\H{o}
polynomials. Indeed, by
introducing the $q-$dilatation operator
 $_{q}T_{y}f(y)=f(qy)$ and
recalling  that the $q-$exponential (\ref{num1})
 obeys the
difference rules
\begin{equation}
_{q}D_{\alpha}e_{q}(\alpha y)=\frac{y}{1-q}
e_{q}(\alpha y)
\end{equation}
one can derive for $\mathcal{S}_{q}(\alpha;y)$
 the following
relation
\begin{equation}
_{q}D_{\alpha}\mathcal{S}_{q}(\alpha;y)=
\frac{1}{1-q}\left(1+y _{q}T^{-1}_{y}\;_{q}
T_{\alpha}\right)
\mathcal{S}_{q}(\alpha;y).
\end{equation}
Also, by applying the operator $_{q}D_{\alpha}$
on the right-hand side
of (\ref{generate}), we arrive at
the relation
\begin{equation}
\label{creation}
{\cal H}_{n+1}(y|q)={\cal H}_{n}(y|q) + yq^{n}
\;_{q}T^{-1}_{y}{\cal H}_{n}(y|q).
\end{equation}
which can be converted into \cite{galetti}
\begin{equation}
{\cal H}_{n+1}(y|q) - (1+y){\cal H}_{n}(y|q)
 + y(1-q^{n}){\cal H}_{n-1}(y|q)=0.
\end{equation}
(\ref{creation}) yields $ {\cal H}_{n+1}(y|q)=
(I +
yq^{N}\;_{q}T^{-1}_{y}){\cal H}_{n}(y|q)$.
Therefore, we state the
following.
\begin{theorem}
Let
\begin{eqnarray}
S_{+}:= I + yq^{N}\;_{q}T^{-1}_{y} \qquad
S_{-}:=\; _{q}D_{y}
\end{eqnarray} be raising and lowering
operators, respectively, and
\begin{eqnarray}
  N_{q}:=S_{+}S_{-}
\end{eqnarray}
Then,   their realizations are performed as:

\begin{eqnarray}
i) \;\; S_{+}{\cal H}_{n}(y|q)&=&
 {\cal H}_{n+1}(y|q)\\
ii)\;\; S_{-}{\cal H}_{n}(y|q)&=& [n]_{q}^{M}
{\cal H}_{n-1}(y|q)
\qquad [n]_{q}^{M}=\frac{1-q^{n}}{1-q}\\
iii) \;\; N_{q} {\cal H}_{n}(y|q)&=&
[n]_{q}^{M}{\cal H}_{n}(y|q).\label{difference}
\end{eqnarray}

\end{theorem}
\begin{corollary}
The following commutation relations hold:

\begin{equation}
i) \;\; [S_{-},S_{+}]=q^{N}\qquad
ii) \;\; [N_{q},S_{+}]=S_{+}q^{N}
\end{equation}
\begin{equation}
iii) \;\;[N,S_{-}]=-S_{-}
\qquad iv) \;\;[N_{q},S_{-}]=-q^{N}S_{-}.
\end{equation}

\end{corollary}

Now, coming back to (\ref{difference}) and
  using the explicit realization
of the raising and lowering operators, i.e.
 $S_{+}$ and $S_{-}$, we can write
\begin{equation}
N_{q}{\cal H}_{n}(y|q)= \left(_{q}D_{y}+
yq^{n}\;_{q}T^{-1}_{y}\;_{q}D_{y}\right)
{\cal H}_{n}(y|q)=[n]_{q}^{M}{\cal H}_{n}(y|q)
\end{equation}
from which, after using the identity
$_{q}T^{-1}_{y}\;_{q}D_{y}
=q_{q}D_{y}\:_{q}T^{-1}_{y}$, we get a
$q-$difference equation obeyed
by the  Rogers-Szeg\H{o}
polynomials
\begin{equation}
\left(_{q}D_{y}+ yq^{n+1}\;_{q}D_{y}
\;_{q}T^{-1}_{y}-[n]_{q}^{M}\right)
{\cal H}_{n}(y|q)=0.
\end{equation}

\section{ $(p,q,\mu,\nu )-$exponential function:
 matrix elements and Fourier-Gauss transform}

As a straightforward generalization of the
 $(q,\mu)-$exponential
function (\ref{gene1}), let us define the
 following
$(p,q,\mu,\nu)-$exponential
\begin{equation}
\label{gene2}
E^{\mu,\nu}_{p,q}(z)=
\sum_{n=0}^{+\infty}\left(\frac{q^{\mu}}{p^{\nu}}
\right)
^{n^{2}}\frac{z^{n}}{[p,q;p,q]_{n}} \qquad
q^{2\mu}p^{1-2\nu}<1 \quad 0<pq<1
\end{equation}
where
\begin{equation}
[p^{\rho},q^{\delta};p,q]_{n}=
\left( \frac{1}{p^{\rho }}-q^{\delta}\right)
\left( \frac{1}{p^{\rho +1}}-q^{\delta+1}\right)
\ldots\left( \frac{1}{p^{\rho +n-1}}-q^{\delta
+n-1}\right)
\end{equation}
with $\rho =\delta =1$ here.
In terms of the $q-$shifted factorial
$(a;q)_{n}$ one has
\begin{equation}
[p^{\rho },q^{\delta};p,q]_{n}=p^{-\left(
n(n-1)/2+\rho  n\right)}(p^{\rho }
q^{\delta};pq)_{n}.
\end{equation}
In the limit $(p,q)\to(1,1)$, once $z$
 has been rescaled by
$(p^{-1}-q)$, all these functions
 tend to the ordinary
 exponential:
\begin{equation}
\lim_{(p,q)\to (1,1)}
E_{p,q}^{\mu,\nu}((p^{-1}-q)z)=\exp(z).
\end{equation}

It is worth noticing that for  $p=1$,
(\ref{gene2}) yields
(\ref{gene1}). For some specific values
 of $\mu$ and $\nu$,
 standard $(p,q)-$exponentials are recovered.
Indeed,  for $\mu=\nu=1/2$  and $\mu=\nu=0$
 one has, respectively,
\begin{eqnarray}
E^{1/2,1/2}_{p,q}(z)&=&E_{p,q}\left(
\left(\frac{q}{p}\right)^{1/2}z\right)\\
E^{0,0}_{p,q}(z)&\equiv&e_{p,q}(z)=
\sum_{n=0}^{+\infty}\frac{z^{n}}{[p,q;p,q]_{n}}
\end{eqnarray}
where $E_{p,q}(z)$ is the $(p,q)-$exponential
 defined by Vinet
{\it et al} \cite{vinet}:
\begin{equation}
 E_{p,q}(z)=
 \sum_{n=0}^{+\infty}\left(\frac{q}{p}\right)
^{n(n-1)/2}\frac{z^{n}}{[p,q;p,q]_{n}}.
\end{equation}

\subsection{
$(p,q,\mu,\nu)-$exponential function:
 matrix elements}

Next, we deal with  the  computation of
 the matrix elements associated with a
generalization of the $(q,\mu)$-exponential
 function (\ref{gene1}).

The $(p,q)-$oscillator algebra is generated by
  three
elements $A_{-}$, $A_{+}$ and $N$ obeying the
 relations
\cite{jang}

\begin{eqnarray}
A_{-}A_{+}-pA_{+}A_{-}=q^{-N}\qquad
A_{-}A_{+}-q^{-1}A_{+}A_{-}=p^{N}
\label{alg1}\\
\;[N,A_{-}]=-A_{-}\qquad [N,A_{+}]=A_{+}.
\label{alg2}
\end{eqnarray}

In general, the two parameters $q$ and $p$
may be real or a
phase factor. Throughout, we suppose that
they are real and positive.

The connection between quantum algebras and
 special functions
 depends  not
only on the particular algebra which is
considered but also on
 the special
realization which is used.
The representation may be useful to
 construct
models for bibasic special functions.

We consider  a realization  of (\ref{alg1})
 and (\ref{alg2})
 in terms of operators acting on the space of
  analytic functions  as
 follows:

\begin{eqnarray}
 A_{-}f(z)&:=&\frac{1}{z(q^{-1}-p)}
 \left[f\left((pq)^{1/2}z\right)-f
 \left((pq)^{-1/2}z\right)
 \right]\\
  A_{+}f(z)&:=&-z(p/q)^{1/2}f
  \left((p/q)^{1/2}z\right)\\
   Nf(z)&:=&z\frac{{\rm d}}{dz}f(z)
\end{eqnarray}

and the basis $\{ \zeta_{n}=z^{n},
 n\in \mathbb{N} \}$
 such that

\begin{eqnarray}
A_{+}\zeta_{n}&=&-\left(\frac{q}{p}
\right)^{-(n+1)/2}
\zeta_{n+1},\\
A_{-}\zeta_{n}&=&\left(\frac{q}{p}
\right)^{1+n/2}
\left(\frac{p^{n}-q^{-n}}{p^{-1}-q}
\right)\zeta_{n-1}.
\end{eqnarray}

For an algebraic interpretation of special
class of
bibasic functions, let us  introduce the
following operators
\begin{equation}
\mathcal{U}^{(\mu,\nu)}(\alpha,\beta)
=E_{p,q}^{\mu,\nu}(\alpha(p^{-1}-q)A_{+})
E_{p,q}^{\mu,\nu}(\frac{\beta p}{q}(p^{-1}-q)
A_{-}).
\end{equation}
 Then,  in the limit
$(p,q)\to(1,1)$,  they go into the Lie group
element
$\exp(\alpha
A_{+})\exp(\beta A_{-}).$ Their matrix elements,
 in the
representation space spanned by the vectors
 $\zeta_{n}$, are defined
by
\begin{equation}
 \mathcal{U}^{(\mu,\nu)}(\alpha,\beta)\zeta_{n}
 =\sum_{m=0}^{+\infty}U_{m,n}^{\mu,\nu}(\alpha,
 \beta)
 \zeta_{m}
\end{equation}
and, when evaluated, are found to involve
 generalized bibasic hypergeometric
 functions.
 Explicitely, after a straightfoward computation,
  one obtains

 \begin{eqnarray}
  U^{(\mu,\nu)}_{m,n}(\alpha,\beta)
   &=&(-\beta)^{n-m}\left[
\begin{array}{l}
n\\
m
\end{array}\right]_{p,q}
\left(\frac{q^{\mu-1/4}}{p^{\nu-1/4}}\right)
^{(m-n)^{2}}
\left(\frac{q}{p}\right)^{-(n-m)(1+2m)/4}
\nonumber \\
 &&\times \mathcal{L}^{(n-m;\mu,\nu)}_{m}
 (-\alpha\beta;p,q)
 \quad {\rm
if}\quad m\leq n
\label{exp1}\\
U^{(\mu,\nu)}_{m,n}(\alpha,\beta)
 &=& \frac{[-\alpha(p^{-1}-q)]^{m-n}}
 {\prod_{l=1}^{m-n}
 (p^{-l}-q^{l})}
 \left(\frac{q^{\mu-1/4}}{p^{\nu-1/4}}
 \right)^{(m-n)^{2}}
\left(\frac{q}{p}\right)^{-(m-n)(1+2n)/4}
\nonumber \\
 &&\times \mathcal{L}^{(m-n;\mu,\nu)}_{n}
 (-\alpha\beta;p,q)
 \quad {\rm
if}\quad m\geq n
\label{exp2}
\end{eqnarray}

where $\mathcal{L}^{(\gamma;\mu,\nu)}_{n}
(x;p,q)$ are the polynomials
given by
\begin{eqnarray}
 \mathcal{L}^{(\gamma;\mu,\nu)}_{n}(x;p,q)
 &=& \sum_{k=0 }^{n}
\left(\frac{q^{\mu}}{p^{\nu}}\right)^{2k
(\gamma+k)}
\frac{((pq)^{-n};pq)_{k}}{(pq;pq)_{k}
((pq)^{\gamma +1};pq)_{k}}
\nonumber \\
&&\times p^{k( k
+1)/2}\left[x(1-pq)p^{\gamma+n}\right]^{k}.
\end{eqnarray}
 Note that in passing from expression
 (\ref{exp1}) to
  (\ref{exp2})  for
 $U^{(\mu,\nu)}_{m,n}(\alpha,\beta)$ or vice
  versa, $m$
   and $n$ are exchanged in the polynomials
$\mathcal{L}^{(\gamma;\mu,\nu)}_{n}(x;p,q)$.

The connection with
standard $(p,q)$-bibasic hypergeometric
 functions is observed
for particular values of $\mu$ and $\nu$.

 For $\mu=\nu=0$
\begin{eqnarray}
\mathcal{L}^{(\gamma;0,0)}_{n}(x;p,q)
= \Phi\left[
\begin{array}{cc}
 (pq)^{-n},0:-&\\
                 &;pq,p,-x(1-pq)p^{\gamma+n+1}\\
  (pq)^{\gamma+1},-:0 &
\end{array}
\right]
\end{eqnarray}
for $\nu=\mu=1/4 $
\begin{eqnarray}
&&\mathcal{L}^{(\gamma;1/4,1/4)}_{n}(x;p,q)
= \nonumber \\
&& \Phi\left[
\begin{array}{cc}
 (pq)^{-n}:0&\\
                 &;pq,p,
x(1-pq)\left(\frac{q}{p}\right)^{(\gamma+1)/2}
p^{\gamma+n+1}\\
(pq)^{\gamma+1}:- &
\end{array} \right]
\end{eqnarray}

where
  \begin{eqnarray}
  \Phi\left[
\begin{array}{cc}
   \underline{a}:\underline{c}&\\
                 &;q,p,z\\
   \underline{b}:\underline{d}&
   \end{array}
 \right] &=&\sum_{l=0}^{+\infty}
   \frac{(\underline{a};q)_{l}
   (\underline{c};p)_{l}}
   {(q;q)_{l}(\underline{b};q)_{l}
   (\underline{d};p)_{l}}
   \left[(-1)^{l}q^{\frac{l(l-1)}{2}}
   \right]^{1+m-n}\nonumber \\
   &&\times \left[(-1)^{l}p^{\frac{l(l-1)}{2}}
   \right]^{s-r}z^{n}
   \label{op4}
   \end{eqnarray}
   is the well-known bibasic hypergeometric
   series
   \cite{gasper}.
In
(\ref{op4})
\begin{eqnarray}
\underline{a}=(a_{1},\ldots,a_{n})&&\quad
\underline{c}=(c_{1},\ldots,c_{r}) \nonumber \\
\underline{b}=(b_{1},\ldots,b_{m})&&\quad
\underline{d}=(d_{1}
\ldots,d_{s})
\end{eqnarray}
\begin{equation}
(\underline{a};q)_{l}=(a_{1};q)_{l}
\ldots(a_{n};q)_{l}.
\end{equation}

\subsection{
$(p,q,\zeta )-$exponential function:
 Fourier-Gauss transform}

\begin{proposition}
Let
 $\mu=\nu= \zeta /2$.  Then, we obtain the
$(p,q,\zeta )-$exponential function
\begin{eqnarray}
\label{exp}
E^{\zeta /2, \zeta /2}_{p,q}(z)&\equiv&
E^{(\zeta )}_{p,q}(z)=\sum_{n=0}^{+\infty}
\left(\frac{q}{p}\right)^{\zeta n^{2}/2}
\frac{z^{n}}{\left[p,q;p,q\right]_{n}}.
\end{eqnarray}
to which  correspond the following  two types
 of Fourier-Gauss transforms
\begin{eqnarray}
E^{(\zeta +1/2)}_{p,q}\left(te^{-kx}\right)
e^{-x^{2}/2}
=\frac{1}{\sqrt{2\pi}}\int_{-\infty}^{+\infty}
e^{ixy-y^{2}/2}E^{(\zeta )}_{p,q}(te^{iky})
{\rm d}y
\label{fou1}
\end{eqnarray}
and
\begin{eqnarray}
E^{(\zeta -1/2)}_{p,q}\left(te^{ikx}\right)
e^{-x^{2}/2}
=\frac{1}{\sqrt{2\pi}}\int_{-\infty}^{+\infty}
e^{ixy-y^{2}/2}E^{(\zeta )}_{p,q}(te^{ky})
{\rm d}y
\label{fou2}
\end{eqnarray}
where $q=p\exp{(-2k^{2})}$.
\end{proposition}

{\bf Proof.}

 To  evaluate the right-hand
 side of
(\ref{fou1}) (resp. (\ref{fou2})) one only
 needs to use the
 definition (\ref{exp})
with $z=te^{iky}$ (resp. $z=te^{ky}$) and
to integrate
the sums termwise
by the Fourier transform
\begin{eqnarray}
\frac{1}{\sqrt{2\pi}}\int_{-\infty}^{+\infty}
e^{ixy-y^{2}/2}{\rm d}y=e^{-x^{2}/2}
\end{eqnarray}
for the Gauss exponential function
 $\exp({-x^{2}/2}).$
$\Box$

Important particular cases of (\ref{fou1}) are
\begin{eqnarray}
\epsilon_{p,q}\left(te^{-kx}\right)e^{-x^{2}/2}
=\frac{1}{\sqrt{2\pi}}\int_{-\infty}^{+\infty}
e^{ixy-y^{2}/2}e_{p,q}(te^{iky}){\rm d}y
\label{fou11}
\end{eqnarray}
and
\begin{eqnarray}
E_{p,q}\left(\left(\frac{q}{p}\right)^{1/2}
te^{-kx}\right)
e^{-x^{2}/2} =\frac{1}{\sqrt{2\pi}}
\int_{-\infty}^{+\infty}
e^{ixy-y^{2}/2}\epsilon_{p,q}(te^{iky})
{\rm d}y
 \label{fou12}
\end{eqnarray}
where
\begin{equation}
 \epsilon_{p,q}(z)=E_{p,q}^{(1/2)}(z).
\end{equation}
These case correspond to the values $0$
and $1/2$ of the
parameter
 $\zeta$, respectively.

When $\zeta =1/2$, from (\ref{fou2}) follows
 the inverse
 Fourier transformation
with respect to (\ref{fou11})
\begin{eqnarray}
e_{p,q}\left(te^{ikx}\right)e^{-x^{2}/2}
=\frac{1}{\sqrt{2\pi}}\int_{-\infty}^{+\infty}
e^{ixy-y^{2}/2}\epsilon_{p,q}(te^{ky}){\rm d}y
\label{fou21}
\end{eqnarray}
whereas the value $\zeta =1$ yields the inverse
to (\ref{fou12}),
 i.e.
\begin{eqnarray}
\epsilon_{p,q}\left(te^{ikx}\right)e^{-x^{2}/2}
=\frac{1}{\sqrt{2\pi}}\int_{-\infty}^{+\infty}
e^{ixy-y^{2}/2}E_{p,q}\left(\left(\frac{q}{p}
\right)^{1/2}te^{ky}
\right){\rm d}y.
\label{fou22}
\end{eqnarray}

Actually, the Fourier-Gauss transforms
(\ref{fou1}) and
 (\ref{fou2}) may be
written in the unified form
\begin{eqnarray}
E_{p,q}^{(\zeta +\varrho^{2}/2)}
 \left(te^{-\varrho kx}\right)
e^{-x^{2}/2}
=\frac{1}{\sqrt{2\pi}}\int_{-\infty}^{+\infty}
e^{ixy-y^{2}/2}E_{p,q}^{(\zeta)}
(te^{i\varrho ky}){\rm d}y.
\label{fouu}
 \end{eqnarray}
This is easy to prove in exactly the same way
 as (\ref{fou1})
 and
(\ref{fou2}). It is worth noticing that, for
 $\zeta =0$ and
$\varrho=\sqrt{2}$,  the Fourier-Gauss transform
 (\ref{fouu}) gives
the following relation
\begin{eqnarray}
 E_{p,q}
\left(\left(\frac{q}{p}\right)^{1/2}
te^{\varrho kx}\right)
e^{-x^{2}/2} =\frac{1}{\sqrt{2\pi}}
\int_{-\infty}^{+\infty}
e^{ixy-y^{2}/2}e_{p,q}(te^{i\varrho ky})
{\rm d}y
 \label{fouu1}
 \end{eqnarray}
  which is a  particular version
of the $(p,q)-$Ramanujan's
integral with respect to
a complex parameter \cite{Raman}.

The formalism  used here provides
  with a straightforward algorithm for
  characterizing
 $q-$polynomials. In this paper, we have
  shown that
Rogers-Szeg\H{o} polynomials also could be
 studied along the same
basis. The key of all these investigations
undoubtedly remains the
expression of conveniently chosen deformed
exponential function. On
this basis, using the relation (\ref{gene2})
 giving a generalized
$(p,q,\mu, \nu)-$exponential function,
this formalism  may also turn
out to be useful for achieving a global
definition as well as a
better understanding of $(p,q)-$analogs of
 special functions. These
aspects are now under consideration.

\section*{Acknowlegments}
This work is partially
supported by the Abdus Salam International
Centre for Theoretical
Physics (ICTP, Trieste, Italy) through the
Office of External
Activities (OEA) - \mbox{Prj-15}. The ICMPA
 is in partnership with
the Daniel Iagolnitzer Foundation (DIF),
France.

\end{document}